\documentclass[twoside]{article}
\usepackage{ijmpal, latexsym, amssymb, kec, graphicx, epsfig}  
\begin{document}
\runninghead{Toward a Neutrino Mass Matrix} 
{Toward a Neutrino Mass Matrix}

\normalsize\textlineskip
\thispagestyle{empty}
\setcounter{page}{1}

\copyrightheading{}			%{Vol. 0, No. 0 (1993) 000--000}

\vspace*{0.88truein}

\fpage{1}
\centerline{\bf TOWARD A NEUTRINO MASS MATRIX}
\vspace*{0.37truein}
\centerline{\footnotesize KEVIN CAHILL}
\vspace*{0.015truein}
\centerline{\footnotesize\it Department of Physics and Astronomy, 
University of New Mexico}
\baselineskip=10pt
\centerline{\footnotesize\it Albuquerque, NM 87131-1156, USA}

%\vspace*{0.225truein}
%\publisher{(received date)}{(revised date)}

\vspace*{0.21truein}
\abstracts{One may identify the general properties
of the neutrino mass matrix by generating many
random mass matrices and testing them 
against the results of the neutrino experiments.}{}{}
\textlineskip			%) USE THIS MEASUREMENT WHEN THERE IS
\vspace*{12pt}			%) NO SECTION HEADING

%\vspace*{1pt}\textlineskip	%) USE THIS MEASUREMENT WHEN THERE IS
%\section{The Neutrino Mass Matrix}	%) A SECTION HEADING
%\vspace*{-0.5pt}
\noindent
There are three light, active neutrinos whose fields 
\(\nu_e\), \(\nu_\mu\), and \(\nu_\tau\) 
are left handed;
there also probably are three right-handed neutrinos
whose fields \( \nu_{re}\), \(\nu_{r\mu}\), and \(\nu_{r\tau}\)
do not participate
in the electroweak interactions and
are said to be \emph{sterile}\@.
Fields that mix must transform in the same way.
So it makes sense to combine the three left-handed
(two-component) 
neutrinos \(\nu_e\), \(\nu_\mu\), and \(\nu_\tau\)
with the left-handed fields \(s_e\), \(s_\mu\), and \(s_\tau\)
that are the charge conjugates
of the right-handed neutrino fields
\( \nu_{re}\), \(\nu_{r\mu}\), \(\nu_{r\tau}\), i.e., 
\( s_e = - i \s_2 \, \nu_{re}^* 
= (\nu_{re2}^\dgr, - \nu_{re1}^\dgr)\), etc\@. 
%Here hermitian conjugation 
%without transposition is denoted by an asterisk, and
%\( \s_2 \) is the second Pauli spin matrix\@.
The six left-handed neutrino fields
form a six-vector, 
\(N = (\nu_e,\nu_\mu,\nu_\tau,s_e,s_\mu,s_\tau)\)\@. 
The only mass term available for 
two left-handed neutrino fields \(N_i\) and \(N_j\) is
\( \mathcal{M}_{ij} \, ( N_{i1} N_{j2} - N_{i2} N_{j1} ) 
+ \, \mathrm{h.c}\)\@. 
% \(m \, (\chi_1 \psi_2 - \chi_1 \psi_2) \) 
%\( i \mathcal{M}_{ij} N_i^\top \sigma_2 N_j 
%- i \mathcal{M}_{ij}^* N_j^\dagger \sigma_2 N_i^*
%+ \, \mathrm{h.c.} \, \) in which
%\(\top\) means transpose. 
Because the fields \(N_i\) anticommute,
the six-by-six complex mass matrix 
\beq
\cM = \pmatrix{ F & D \cr
                D^\top & E\cr}
\label {cM}
\eeq
is symmetric; here \(\top\) means transpose.  
In most models the quantities 
\( \mathcal{M}_{ij} \) are the mean values 
of Higgs fields in the vacuum. 
Because the complex mass matrix 
\( \mathcal{M} \) is symmetric,
it may be factored
\(\mathcal{M} = U \, M \, U^\top\)
by a matrix \(U\) that is unitary,
and a matrix \(M\) that is diagonal and positive.$^1$ 
The elements \(m_i\) of \(M\) are the masses of the 
neutrinos.  The vector \(N_m\) of six mass eigenfields 
is \(N_m = U^\top N \)\@.
The mass eigenfields %\(\nu_{m_i}\)
%\beq
%\nu_{m_i} = \sum_{i=1}^6 U_{ji} N_j
%\eeq 
are the normal modes of the action. 
They are Majorana neutrinos.
\par
%\section{A Statistical Approach}
The six-by-six complex mass matrix \( \mathcal{M} \) 
involves 42 real parameters. 
One may explore this large space 
and test various 
properties of the mass matrix \(\cM\)
by generating many 
random mass matrices that share those
properties,
and by computing the extent to which they 
fit the results of the neutrino experiments. 
For instance, one may define an angle \(x_\nu\) by
\(
\sin^2 x_\nu = \mathrm{Tr}( E^\dgr E + F^\dgr F ) /
\mathrm{Tr}( 2 D^\dgr D + E^\dgr E + F^\dgr F )
%\label {x}
\)
and test whether \(x_\nu\) can be very small.$^2$
This angle characterizes where 
the six neutrinos lie on a continuum
that extends from three purely Dirac neutrinos, \(x_\nu=0\),  
to six purely Majorana neutrinos, \(x_\nu=\pi/2\). 
When the angle \(x_\nu\) is small, 
as in a theory in which \(B-L\)
is nearly conserved, 
the six masses \(m_i\) coalesce into
three pairs of nearly degenerate masses, and the
six neutrinos form three nearly Dirac neutrinos, 
a condition that has been called 
\emph{pseudo-Dirac}.$^3$
This \(B-L\) or small-\(x_\nu\) 
property explains the large mixing angles
and tiny mass differences
seen experimentally and allows the masses of the
neutrinos to lie in the eV range with
\(\sum_i m_i \lsim 8 \) eV
which is a cosmological bound.$^4$ 
To test this property,
I used the software package LAPACK\,$^5$ 
to factorize 10,000 random mass matrices \(\cM\)
every parameter of each of which was a complex number
\( z = x + i y \) with \(x\) and \(y\) chosen
randomly and uniformly from the interval \( [ - 1 \mathrm{eV},
1 \mathrm{eV} ] \)\@.
The solar neutrinos were taken to have an energy of 1~MeV,
and the probability 
\(P(\nu_i \to \nu_{i'}) = |A(\nu_i \to \nu_{i'})|^2\)
with \(A(\nu_i \to \nu_{i'}) 
= \sum_j U^*_{i'j} U_{ij} \exp(-im_j^2L/(2E))\) 
%\beq
%P(\nu_i \to \nu_{i'}) = 
%\sum_{j,j'\,\mathrm{light}}
%U^*_{i'j} U_{ij} U_{i'j'} U^*_{ij'} 
%\exp{\left[i(m_{j'}^2 - m_{j}^2)L/(2E)\right]} 
%\label {Pnunu'}
%\eeq
was averaged 
over one revolution of the Earth about the Sun.
The atmospheric neutrinos were averaged over the atmosphere
and over energies in the range of 1--30 GeV
weighted by the flux of atmospheric muon neutrinos
as a function of energy and local zenith angle
given by the Bartol group.$^6$  
The resulting scatter plots fit the gross features of the
solar and atmospheric experiments quite well
with \( \sin x_\nu = 0.003 \) when two additional properties
are added.$^2$  The first is a constraint on inter-generational mixing
which I imposed by suppressing the singly off-diagonal matrix elements 
of \(D, E,\) and \(F\) by 0.2 and the doubly off-diagonal matrix elements
by \(0.04\)\@.  The second is a weak mass hierarchy which I implemented  
by scaling the \(i,j\)-th elements
of the sub-matrices \( E, F,\) and \(D\) of the mass matrix \(\cM\),
Eq.~(\ref{cM}), 
by the factors \(f(i)*f(j)\) where \( \vec f = ( 0.2, 1, 2 ) \)\@.  
%as shown by Fig.~\ref{fig:h5}.
%\goodbreak
%\begin{figure}[h]  
%\centering
%\resizebox{\textwidth}{!}{
%\includegraphics {h5}    } 
%\includegraphics [0,0][400,310]{h5}
%\caption{The probabilities \( P_{\mathrm{sol}}(\nu_e \to \nu_e ) \)
%and \( P_{\mathrm{atm}}(\nu_\mu \to \nu_\mu ) \)
%for 10,000 random mass matrices \(\cM\)
%all with the parameter \(\sin x_\nu = 0.003\),
%with inter-generational mixing suppressed
%by factors of 0.2, and
%with a weak mass hierarchy.}
%\label {fig:h5}
%\end{figure} 
These three properties also lead to general agreement
with the results of the LSND and KARMEN2 experiments. 
If the physical mass matrix \(\cM\) shares these
properties, 
%In Fig.~\ref{fig:bo}
%for a set of 10,000 mass matrices generated randomly
%with the same properties as in Figs.~\ref{fig:h5},
%the appearance probabilities \(P(\bar\nu_\mu \to \bar\nu_e)\)
%and \(P(\bar\nu_\mu \to \bar\nu_\mu)\) for
%MiniBooNE are plotted.
and if MiniBooNE can achieve a sensitivity
of 0.001 for \(\bar\nu_\mu \to \bar\nu_e\)
and a precision of 0.01 for \(\bar\nu_\mu \to \bar\nu_\mu\),
then it has a good chance of seeing
the appearance of \(\bar\nu_e\)
and the disappearance of \(\bar\nu_\mu\)\@.$^2$ 
\par
\begin{figure}[h] 
\centering 
\resizebox{\textwidth}{!}{
\includegraphics{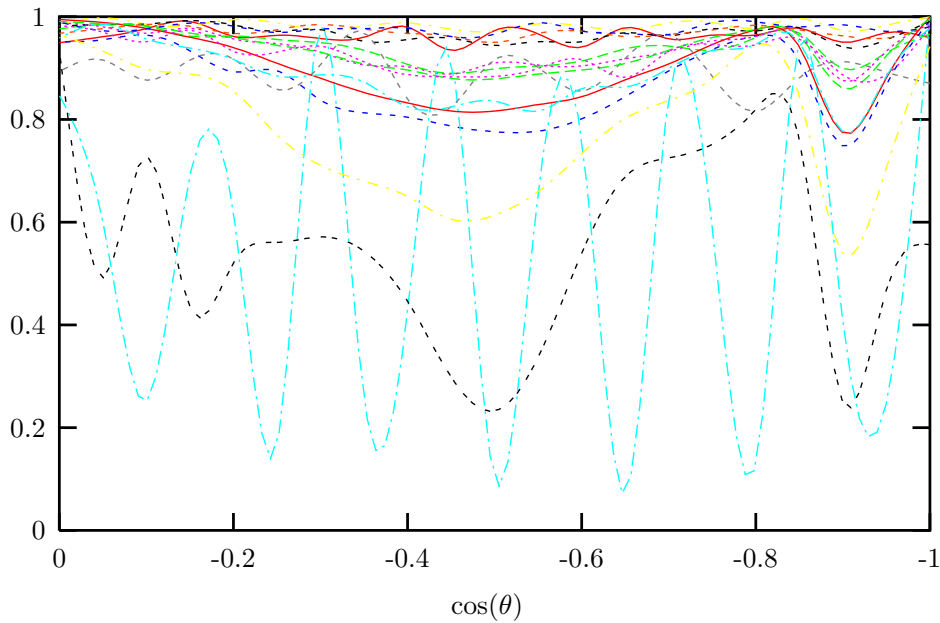}  }
\caption{\small The probability 
that a 100-GeV muon neutrino would traverse the Earth 
are plotted against the cosine of the azimuthal angle \( \theta \) 
for 16 random mass matrices \(\cM\) with
\(\sin x_\nu = 0.003\), little inter-generational mixing, 
and a weak mass hierarchy.} 
\label{fig:dpf}  
\end{figure}

 
Another test of the properties
of the neutrino mass matrix \(\cM\)
is provided by the passage of high-energy neutrinos
through the Earth.$^7$\ %~\cite{MSW,HAB,NR}  
In momentum space the Dirac equation 
for the 24-vector \( (N,i \sigma_2 N^* )\)
of neutrino fields is
\beq
0=\pmatrix{E + \vec \sigma \cdot \vec p + V & - \cM^\dagger \cr
 - \cM & E - \vec \sigma \cdot \vec p - V \cr }
\pmatrix{ N \cr i \sigma_2 N^* \cr } 
\label {24eq}
\eeq 
in which the hermitian matrix \( V \) represents
the interaction of the neutrinos with the electrons
and quarks of the Earth.  The matrix \( V \) 
is diagonal with elements
\( V = (G_F/\sqrt{2}) \, 
(2 N_e - N_n, - N_n, - N_n, 0, 0, 0 ) \)
in which \(N_e\) and \(N_n\) are the 
number densities of electrons and neutrons.
%If sterile neutrinos interact with quarks
%and leptons via heavy gauge bosons, then 
%one should alter the three
%zero elements of the matrix \(V\) to reflect that.
By neglecting neutrinos of positive helicity, 
antineutrinos of negative helicity,
and the terms \(V^2\) and \([V,\cM]\),
one finds for the square of the mass operator
the six-by-six hermitian matrix
\(
M(p)_{\pm}^2 = \cM^\dagger \cM \, \pm \, 2 \, |\vec p| \, V 
= U^\dagger \, M_d(p)^2_+ \, U
%\label{M2} 
\)
in which \(U\) is another unitary matrix
and the plus sign is for neutrinos
and the minus sign for anti-neutrinos. 
I used the PREM %~\cite{PREM}
model$^8$ to divide the Earth into 82 shells each with its
approximate values of the densities \(N_e\) and \(N_n\)\@. 
My code computes the amplitude for a neutrino 
to propagate through the Earth, which is the path-ordered
exponential
\beq
A = \mathcal{P} \left\{ 
\exp\left[ {i \int \! dx \, M(p)^2_+/(2 |\vec p|)} 
\right] \right\},
\eeq 
as a product of up to 82 factors of the form
\(U^\dagger \exp\left[ 
{i \, \Delta x \, M_d(p)^2_+/(2 |\vec p|)} \right] U \)\@. 
For 16 random mass matrices \(\cM\) with 
small \(x_\nu\),
little inter-generational mixing, and a weak mass
hierarchy, the probabilities 
\(P_{\mathrm{SK}}(\nu_\mu \to\nu_\mu)\) 
that a muon neutrino of energy
100 GeV would traverse the Earth
at a given azimuthal angle \(\theta\)
and enter the Super Kamiokande detector
are plotted in 
Fig.~\ref{fig:dpf}. 
\noindent 
Because high-energy 
interactions with matter suppress 
active-sterile neutrino oscillations,
the probability \(P_{\mathrm{SK}}(\nu_\mu \to\nu_\mu)\)
remains close to unity in 13 of the cases.
But in three cases it 
drops quite far below unity.
Further work is required to determine what additional properties,
if any, would cause such random mass matrices to display
the oscillations and neutral-current events observed 
by the Super-Kamiokande collaboration.$^9$   
\textheight=7.8truein
\setcounter{footnote}{0}
\renewcommand{\thefootnote}{\alph{footnote}}
%\nonumsection{Acknowledgements}
%\noindent
%This section should come before the References. 

\nonumsection{References}
\noindent
%References are to be listed in the order cited in the text. Use
%the style shown in the following examples. For journal names,
%use the standard abbreviations. Typeset references in 9 pt Times
%Roman.

\end{document}